\documentclass[12pt]{article}

\usepackage{graphicx}
\begin{document}

\newcommand{\siml}{\stackrel{<}{\sim}}
\newcommand{\simg}{\stackrel{>}{\sim}}
\newcommand{\lleq}{\stackrel{<}{=}}

\baselineskip=1.333\baselineskip


%
\begin{center}
{\large\bf
Stationary and dynamical properties of finite $N$-unit 
Langevin models subjected to
multiplicative noises 
} 
\end{center}

\begin{center}
Hideo Hasegawa
\footnote{E-mail address:  hasegawa@u-gakugei.ac.jp}
\end{center}

\begin{center}
{\it Department of Physics, Tokyo Gakugei University  \\
Koganei, Tokyo 184-8501, Japan}
\end{center}
\begin{center}
({\today})
\end{center}
\thispagestyle{myheadings}

\begin{abstract}
We have studied
the finite $N$-unit Langevin model 
subjected to multiplicative noises,
by using the augmented moment method (AMM),
as a continuation of our previous paper
[H. Hasegawa, J. Phys. Soc. Jpn. {\bf 75} (2006) 033001].
Effects of couplings on stationary and dynamical properties
of the model have been investigated.
The difference and similarity between the results
of diffusive and sigmoid couplings are studied in details.
Time dependences of average and fluctuations in local and 
global variables calculated by the AMM are
in good agreement with those of direct simulations (DSs).
We also discuss
stationary distributions of local and global variables
with the use of the Fokker-Planck equation (FPE) method and DSs.
It is demonstrated that stationary distributions
show much variety when multiplicative noise
and external inputs are taken into account.

\end{abstract}

\vspace{0.5cm}


\newpage
\section{INTRODUCTION}

The Langevin equation has been widely employed as 
a useful model for a wide range of stochastic phenomena.
Much study has been made on the Langevin model 
for a single unit as well as coupled systems 
(for a recent review, see Ref. [1]).
The Langevin equation has been commonly solved by using the
Fokker-Planck equation (FPE) method \cite{Risken96}.  
For $N$-unit Langevin equations, 
the FPE method leads to $(N+1)$-dimensional
partial equations to be solved with proper
boundary conditions, which is usually very difficult.
Direct simulation (DS) requires the computational time
which grows as $N^2$ with increasing $N$.
As a useful semi-analytical method for
stochastic equations,  Rodriguez and Tuckwell \cite{Rod96}
proposed the moment method in which the first and second
moments of variables are taken into account.
In this approach, original $N$-dimensional
Langevin equations are transformed to $(N/2)(N+3)$-dimensional
deterministic equations.
For example, this figure becomes 65 and 5150
for $N=10$ and $N=100$, respectively.
Based on a macroscopic point of view,
Hasegawa \cite{Hasegawa03a} has proposed 
the augmented moment method (AMM),
in which the dynamics of coupled Langevin
equations is described by a fairly small number ({\it three})
of quantities:
averages and fluctuations of local and global variables.
The AMM has been successfully
applied to a study on the dynamics of coupled stochastic systems 
described by Langevin, FitzHugh-Nagumo and
Hodgkin-Huxley models subjected to additive noises
with global, local or small-world couplings
(with and without transmission delays)
\cite{Hasegawa03b}-\cite{Hasegawa05a}. 
The AMM was originally
developed by expanding variables
around their mean values in the stochastic model
in order to obtain
the second-order moments both for
local and global variables 
\cite{Hasegawa03a}.  
In a recent paper \cite{Hasegawa06},
we have reformulated the AMM with the use of the FPE,
in order to apply the AMM to 
coupled Langevin model subjected
to multiplicative noises, in which
the difficulty of the Ito versus Stratonovich
representations is inherent.

In recent years,
much attention has been paid to
multiplicative noises in addition to additive noises
(for a review of study on multiplicative noises, 
see Ref. \cite{Munoz04}, related references therein).
The stationary distribution of the Langevin model
subjected to multiplicative noises has been considerably
investigated in various contexts
\cite{Munoz04}-\cite{Ante05}.
Interesting phenomena caused by the two noises have
been intensively studied.
It has been realized that
the property of multiplicative noises
is different from that of additive noises
in some respects.
(1) Multiplicative noises induce the phase transition,
creating an ordered state, while additive noises
are against the ordering \cite{Broeck94}-\cite{Munoz05}.
(2) Although the probability distribution in stochastic systems
subjected to additive Gaussian noise follows the Gaussian,
multiplicative Gaussian noises generally yield non-Gaussian distribution
\cite{Schenzle79}-\cite{Ante05}\cite{Wilk00,Hasegawa05b}.
(3) The scaling relation of the effective 
strength for additive noise given by
$\beta(N)=\beta(1)/\sqrt{N}$ is not applicable to
that for multiplicative noise:
$\alpha(N) \neq \alpha(1)/\sqrt{N}$, where $\alpha(N)$ and $\beta(N)$
denote effective strengths of multiplicative
and additive noises, respectively, in the $N$-unit system
\cite{Hasegawa06}.
A naive approximation
of the scaling relation for multiplicative noise: 
$\alpha(N)=\alpha(1)/\sqrt{N}$
as adopted in Ref. \cite{Munoz05} yields the result
which does not agree with that of DSs.

The purpose of 
the present paper is to discuss effects of couplings
on stationary and dynamical properties
of the $N$-unit Langevin model with multiplicative noises,
which has been not investigated in \cite{Hasegawa06}.
The paper is organized as follows.
In Section 2, the AMM is employed for a discussion
on the finite-$N$ Langevin model 
which is subjected to additive 
and multiplicative noises and which is
coupled by diffusive and sigmoid couplings.
Numerical results are presented in Section 3.
Section 4 is devoted to 
discussion and conclusion, where 
the stationary distribution of local and global variables
are studied with the FPE and DS.

\section{AUGMENTED MOMENT METHOD}
\subsection{A Generalized Langevin model}

We have adopted the finite $N$-unit 
Langevin model given by
\begin{eqnarray}
\frac{dx_i}{dt}\!\!&=&\!\!F(x_i) + \alpha G(x_i) \eta_i(t)
+ \beta \xi_i(t)+I_i^{(c)}(t)+I^{(e)}(t), 
\end{eqnarray}
with 
\begin{equation}
I_i^{(c)}(t)=\frac{J}{Z} 
\sum_{k(\neq i)} [x_k(t)-x_i(t)]
+ \frac{K}{Z} \sum_{k(\neq i)}H(x_k(t)),
\hspace{1.0cm}\mbox{($i=1-N$)} 
\end{equation}
and
\begin{eqnarray}
H(x)=\frac{x}{\sqrt{x^2+1}}.
\end{eqnarray}
Here $F(x)$ and $G(x)$ denote arbitrary functions of $x$:
$J$ and $K$ express the diffusive and sigmoid couplings, respectively,
whose effects will be separately discussed in Sections 2.2 and 2.3:
$Z$ $(=N-1)$ stands for the coordination number: 
$I^{(e)}(t)$ is an external input:
$\alpha$ and $\beta$ denote the strengths of multiplicative
and additive noises, respectively, and
$\eta_i(t)$ and $\xi_i(t)$ express zero-mean Gaussian white
noises with correlations given by
\begin{eqnarray}
\langle \eta_i(t)\:\eta_j(t') \rangle
&=& \delta_{ij} \delta(t-t'),\\
\langle \xi_i(t)\:\xi_j(t') \rangle 
&=& \delta_{ij} \delta(t-t'),\\
\langle \eta_i(t)\:\xi_j(t') \rangle &=& 0.
\end{eqnarray}
Although various types of sigmoid functions 
such as ${\rm tanh}(x)$ and $1/[1+\exp(x)]$, {\it etc.}
have been employed in the literature,
we here adopt a simple analytical expression given 
by Eq. (3) \cite{Monteiro02}.

The Fokker-Planck equation for the distribution of
$\hat{p}(\{ x_i \},t)$ 
is given by 
\cite{Haken83}
\begin{eqnarray}
\frac{\partial}{\partial t}\: \hat{p}(\{ x_i \},t) 
&=&-\sum_k \frac{\partial}{\partial x_k}\{ [F(x_k) 
+\frac{\phi\alpha^2}{2}G'(x_k)G(x_k) 
+ I_k] \:\hat{p}(\{ x_i \},t)\}  
\nonumber \\
&+&\frac{1}{2}\sum_{k }\frac{\partial^2}{\partial x_k^2} 
\{[\alpha^2 G(x_k)^2+\beta^2]\:\hat{p}(\{ x_i \},t) \},
\end{eqnarray}
where $I_k=I_k^{(c)}+I^{(e)}$, $G'(x)=dG(x)/dx$, and
$\phi=1$ and 0 in the Stratonovich and Ito representations,
respectively.
The averaged, global variable $X(t)$
is given by
\begin{equation}
X(t)=\frac{1}{N} \sum_i x_i(t),
\end{equation}
for which the Fokker-Planck equation $P(X,t)$ is formally given by
\begin{equation}
P(X,t) = \int \cdots \int \Pi_i \:dx_i \:\hat{p}(\{x_i \},t)
\: \delta\Bigl(X-\frac{1}{N}\sum_i x_i\Bigr).
\end{equation}

We will discuss the property 
of the coupled Langevin model
with the use of the AMM, which is the second-moment
theory for local and global variables \cite{Hasegawa03a,Hasegawa06}.
The moments of local and global variables are defined by 
\begin{eqnarray}
\langle x_i^k \rangle &=& 
\int \Pi_i \:dx_i \: \hat{p}(\{x_i \},t) \:x_i^k, \\
\langle X^k \rangle&=& \int d\:X P(X, t) \:X^k. 
\hspace{1cm}\mbox{($k=1,2,\cdot \cdot$)}
\end{eqnarray}
From Eqs. (1), (7), (8), (10) and (11), equations of motions
for mean, variance and covariance of local variable ($x_i$) 
and global variable ($X$)
are given by \cite{Hasegawa06}
\begin{eqnarray}
\frac{d \langle x_i \rangle}{dt}
&=& \langle F(x_i) \rangle
+\langle I_i \rangle
+\frac{\phi \:\alpha^2}{2} \langle G'(x_i)G(x_i) \rangle, \\
\frac{d \langle x_i \:x_j \rangle}{dt}
&=& \langle x_i\:F(x_j) \rangle 
+ \langle x_j\: F(x_i) \rangle 
+ \langle x_i I_j \rangle + \langle x_j I_i \rangle
\nonumber \\
&+& \frac{\phi\:\alpha^2}{2}
[\langle x_i G'(x_j) G(x_j) \rangle
+ \langle x_j G'(x_i) G(x_i)\rangle] \nonumber \\ 
&+&[\alpha^2\:\langle G(x_i)^2 \rangle +\beta^2]\:\delta_{ij},\\
\frac{d \langle X \rangle}{dt}
&=&\frac{1}{N} \sum_i \frac{d \langle x_i \rangle}{dt}, \\
\frac{d \langle X^2 \rangle}{dt} 
&=& \frac{1}{N^2}\sum_i \sum_j 
\frac{d \langle x_i\:x_j \rangle}{dt},
\end{eqnarray}
where $I_i=I_i^{(c)}+I^{(e)}$.
Equation (12) is adopted in
the mean-field approximation \cite{Broeck97}.
In Ref. \cite{Kawai04},
Eqs. (12) and (13) are employed for a discussion
on the fluctuation-induced phase transition 
in infinite-$N$ stochastic systems.
Equations (14) and (15) play a crucial role
in discussing finite-$N$ systems, 
as will be shown shortly.

In the AMM \cite{Hasegawa03a,Hasegawa06},
we take into account the three quantities:
$\mu$, $\gamma$ and $\rho$
expressing the mean of $X$,
the averaged fluctuations in local variables ($x_i$) and
fluctuations in global variable ($X$), respectively,
which are defined by
\begin{eqnarray}
\mu &=& \langle X \rangle 
= \frac{1}{N} \sum_i \langle x_i \rangle, \\
%
\gamma &=& \frac{1}{N} \sum_i \langle (x_i-\mu)^2 \rangle, \\
%
\rho &=& \langle (X-\mu)^2 \rangle.
\end{eqnarray}
Expanding $x_i$ in Eqs. (12)-(15) around the average value 
of $\mu$ as
\begin{equation}
x_i=\mu+\delta x_i,
\end{equation}
and retaining up to the order of $ \langle \delta x_i \delta x_j \rangle$, 
we get
equations of motions for $\mu$, $\gamma$ and $\rho$ given by 
\begin{eqnarray}
\frac{d \mu}{dt}&=& f_0+f_2\gamma  
+ K [h_0+h_2 \gamma] 
+\left( \frac{\phi \: \alpha^2}{2}\right)
[g_0g_1+3(g_1g_2+g_0g_3)\gamma]+I^{(e)}, \\
\frac{d \gamma}{dt} &=& 2f_1 \gamma 
+ \left( \frac{2 J N}{Z} \right)(\rho-\gamma)
+\left( \frac{2 K h_1 N}{Z} \right)
\left( \rho-\frac{\gamma}{N} \right) \nonumber \\
&+& (\phi+1) (g_1^2+2 g_0g_2)\alpha^2\gamma 
+ \alpha^2 g_0^2+\beta^2, \\
\frac{d \rho}{dt} &=& 2 f_1 \rho+ 2 K h_1 \rho 
+(\phi+1)(g_1^2+2 g_0g_2)\alpha^2\rho 
+ \frac{\alpha^2 g_0^2}{N}  + \frac{\beta^2}{N},
\end{eqnarray}
where $f_{\ell}=(1/\ell !)
\partial^{\ell} F(\mu)/\partial x^{\ell}$, 
$g_{\ell}=(1/\ell !) 
\partial^{\ell} G(\mu)/\partial x^{\ell}$, and
$h_{\ell}=(1/\ell !)
\partial^{\ell} H(\mu)/\partial x^{\ell}$.
Original $N$-dimensional stochastic equations 
given by Eqs. (1)-(3)
are transformed to three-dimensional deterministic equations
given by Eqs. (20)-(22). 
The stability of Eqs. (20)-(22) may be examined by calculating
their Jacobian matrix, as will be discussed shortly. 
We note that equations of motions for $\mu$ and $\rho$
in Eqs. (20) and (22)
do not include the term of $J$ for diffusive coupling,
while they include the term of $K$ for the sigmoid coupling.
This is because the average of
$\sum_i \langle I_i \rangle$ and
$\sum_{i j} \langle x_i I_j \rangle$ in Eqs. (12) and (15) vanish
due to the nature of the diffusive coupling given in Eq. (2). 
If we consider the conditional average of
$\langle x_i \rangle_i$ for a given site $i$,
its equation of motion has a term relevant to the coupling $J$,
as discussed in Ref. \cite{Birner02}. 

\subsection{Diffusive couplings}

For the linear Langevin model given by
$F(x)=-\lambda x$ and $G(x)=x$
with diffusive couplings ($J \neq 0$, $K=0$),
we get equations of motions for $\mu$, $\gamma$ 
and $\rho$ in the Stratonovich representation ($\phi=1$)
given by
\begin{eqnarray}
\frac{d \mu}{dt}&=&-\lambda \mu 
+ \frac{\alpha^2 \mu}{2}+I^{(e)}, \\
\frac{d \gamma}{dt} &=& -2 \lambda \gamma 
+ \left( \frac{2 J N}{Z} \right)(\rho-\gamma) + 2 \alpha^2 \gamma 
+ \alpha^2 \mu^2 + \beta^2, \\
\frac{d \rho}{dt} &=& -2\lambda \rho
+ 2 \alpha^2 \rho
+ \frac{\alpha^2 \mu^2}{N}  + \frac{\beta^2}{N}.
\end{eqnarray}
The stability of the stationary solutions given 
by Eqs. (23)-(25)
may be examined by calculating their Jacobian matrix.
We get three eigenvalues of
$\lambda -\alpha^2/2$,
$2 \lambda -2 \alpha^2+ 2JN/Z$ and
$2 \lambda - 2 \alpha^2$, from which
the stability condition of the
stationary solution is given by $\alpha^2 < \lambda$.
The stable stationary solutions for $I^{(e)}=I$ are given by
\begin{eqnarray}
\mu&=&\frac{I}{(\lambda-\alpha^2/2)}, \\
\gamma &=& \frac{(\alpha^2 \mu^2+\beta^2)[1+J/Z(\lambda-\alpha^2)]}
{2(\lambda-\alpha^2 + JN/Z)}, \\
&\rightarrow& \frac{(\alpha^2\mu^2+\beta^2)}{2(\lambda-\alpha^2+J)}, 
\hspace{0.5cm}\mbox{(as $N \rightarrow \infty$)}\\
\rho &=& \frac{(\alpha^2 \mu^2+\beta^2)}
{2N(\lambda-\alpha^2)}, \\ 
\frac{\rho}{\gamma} &=& 
\frac{1}{N}
\left(\frac{\lambda-\alpha^2 + JN/Z}{\lambda -\alpha^2 + J/Z} \right), \\
&\rightarrow& \frac{1}{N}
\left( \frac{\lambda-\alpha^2 + J}{\lambda-\alpha^2} \right). 
\hspace{0.5cm}\mbox{(as $N \rightarrow \infty$)}
\end{eqnarray}

\subsection{Sigmoid couplings}

We will make 
an analysis of the linear Langevin model with 
$F(x)=-\lambda x$ and $G(x)= x$
for sigmoid couplings ($J=0$, $K \neq 0$) in Eqs. (2) and (3), 
for which equations of motion for $\mu$, $\gamma$ and $\rho$
in the Stratonovich representation are given by
\begin{eqnarray}
\frac{d \mu}{dt}&=&-\lambda \mu 
+ \frac{\alpha^2 \mu}{2}+K (h_0+h_2 \gamma) +I^{(e)}, \\
\frac{d \gamma}{dt} &=& -2 \lambda \gamma 
+ \left( \frac{2 K h_1 N}{Z} \right)
\left( \rho-\frac{\gamma}{N} \right) 
+ 2 \alpha^2 \gamma 
+ \alpha^2 \mu^2 + \beta^2, \\
\frac{d \rho}{dt} &=& -2\lambda \rho
+ 2 \alpha^2 \rho + 2 K h_1 \rho
+ \frac{\alpha^2 \mu^2}{N}  + \frac{\beta^2}{N},
\end{eqnarray}
where $h_0=\mu/\sqrt{\mu^2+1}$, 
$h_1=1/(\mu^2+1)^{3/2}$ and $h_2=-(3\mu/2)/(\mu^2+1)^{5/2}$.
The stationary solutions with $I^{(e)}=I$ for a small $\mu$
for which $H(\mu) \sim \mu$ are given by
\begin{eqnarray}
\mu&=&\frac{I}{(\lambda-\alpha^2/2-K)}, \\
\gamma &=& \frac{(\alpha^2 \mu^2+\beta^2)[1+K/Z(\lambda-\alpha^2-K)]}
{2(\lambda-\alpha^2+K/Z)}, \\
&\rightarrow& \frac{(\alpha^2\mu^2+\beta^2)}{2(\lambda-\alpha^2)}, 
\hspace{0.5cm}\mbox{(as $N \rightarrow \infty$)} \\
\rho &=& \frac{(\alpha^2 \mu^2+\beta^2)}
{2N(\lambda-\alpha^2-K)}, \\ 
\frac{\rho}{\gamma} &=& 
\frac{1}{N}
\left(\frac{\lambda-\alpha^2+K/Z}{\lambda -\alpha^2-K + K/Z} \right), \\
&\rightarrow& \frac{1}{N}
\left( \frac{\lambda-\alpha^2}{\lambda-\alpha^2-K} \right). 
\hspace{0.5cm}\mbox{(as $N \rightarrow \infty$)}
\end{eqnarray}
Three eigenvalues of Jacobian matrix 
relevant to Eqs. (35), (36) and (38) are
$\lambda-\alpha^2/2-K$, $\lambda-\alpha^2-K$ and
$\lambda-\alpha^2+K/Z$, from which we get
the stability condition of $\alpha^2 < \lambda-K$.

\subsection{Comparison between diffusive and sigmoid couplings}

Comparing Eqs. (23)-(25) for the diffusive coupling
with Eqs. (32)-(34) for the sigmoid coupling,
we note the following difference and similarity in 
$\mu$, $\gamma$ and $\rho$.

\noindent
(i) $\mu$ for the diffusive coupling is independent of 
the coupling ($J$) while
$\mu$ for the sigmoid coupling depends on its coupling ($K$).

\noindent
(ii) When the (positive) coupling is introduced, 
$\gamma$ for the diffusive coupling
is decreased while $\gamma$ for the sigmoid coupling 
is almost independent of it because $\rho \sim \gamma/N$
for small $K$ in the second term of Eq. (33).

\noindent
(iii) When the (positive) coupling is introduced, 
$\rho$ for the sigmoid coupling is increased while
$\rho$ for the diffusive coupling is independent of it.

\noindent
(iv) With increasing the (positive) coupling strength,
the ratio of $\rho/\gamma$ is increased for both the couplings.
This leads to an increased synchronization ratio of $S(t)$:
\begin{equation}
S(t)=\left( \frac{\;\rho(t)/\gamma(t)-1/N}{1-1/N} \right)
=\left( \frac{N \rho(t)-\gamma(t)}{(N-1) \gamma(t)} \right),
\end{equation}
which is one and zero for the completely synchronous
and asynchronous states, respectively \cite{Hasegawa03a}\cite{Note1}.

\subsection{Nature of the AMM}

Before proceeding to the next section of numerical results,
we will discuss the nature of the AMM,
which is essentially the second-moment 
approximation for local and global variables.
One of disadvantages of the AMM is that its applicability
is limited to the weak-noise case
because higher-order moments are assumed to be neglected.
The second-moment given by Eq. (10)
is positive definite for 
the magnitude of the multiplicative noise $\alpha$, as given by
\begin{eqnarray}
\langle x^2 \rangle = \frac{\beta^2}{2(\lambda-\alpha^2)}
< + \:\infty,
\hspace{1cm} \mbox{for $\alpha^2 < \lambda$}
\end{eqnarray}
in the case of $I=J=K=0$.
A simple calculation leads to
the equation of motion of the $k$-th moment 
for even $k$ given by
\begin{eqnarray}
\frac{\partial \langle x^k \rangle}{\partial t} 
&=& - \left( k \lambda -\frac{k^2 \alpha^2}{2} \right) 
\langle x^k \rangle
+ \frac{k (k-1) \beta^2}{2} \langle x^{k-2} \rangle,
\hspace{1cm}\mbox{($k=2,4, \cdot \cdot$)}
\end{eqnarray}
from which the stationary value of $ \langle x^{k} \rangle$ is
given by
\begin{eqnarray}
\langle x^{k} \rangle 
&=& \frac{(k-1)\beta^2}{\; 2 (\lambda - k \alpha^2/2)} \:
\langle x^{k-2} \rangle, \\ 
&=& \frac{(k-1)!! \: \beta^{k}}
{\;2^{k/2} \:\Pi_{\ell=1}^{k/2} (\lambda - \ell \: \alpha^2)}.
\end{eqnarray}
We get the positive definite $\langle x^{k} \rangle$
for $\alpha^2 < 2 \lambda/k$, which
implies that for $2 \lambda/k < \alpha^2 < \lambda$
with $k \geq 4$,
the $k$-th moment may diverge
even if $\langle x^2 \rangle$ remains finite. This 
might throw some doubt on the validity of the AMM
for the multiplicative noise. 
Equations (43) expresses that the motion of 
$\langle x^{k} \rangle$ depends on those of
its lower moments ($\leq k-2$), 
but does not on its higher moments ($\geq k+2$).
Even if $\langle x^{4} \rangle$ diverges, for example,
it has no effects on the motion of $\langle x^{2} \rangle$.
We hope that our AMM is meaningful and useful
for discussions on stochastic systems subjected to
multiplicative noise,
because the results of the AMM
are in good agreement with those of DS,
as will be demonstrated in our numerical calculations.
The advantage of the AMM is that we can easily
discuss the dynamics of $N$-unit Langevin model by
solving the three-dimensional ordinary differential equations.
Note that
it is much more tedious to solve $(N+1)$-dimensional partial
differential equations in FPE and $N$-dimensional 
stochastic equations in DS.

\section{Numerical results}

\subsection{Stationary property}

\subsubsection{Size ($N$) dependence}

We have performed numerical calculations
for linear Langevin models,
solving AMM equations
by the Runge-Kutta method with a time step of 0.01.
Direct simulations for the $N$-unit Langevin model
have been performed by using
the Heun method with a time step of 0.0001.
Results shown in the paper
are averages of 1000 trials otherwise noticed.

The $N$-dependences of
$\gamma$ and $\rho$ in the stationary states
for the diffusive couplings
of $J=0.0$, 0.2 and 0.5 are plotted in Fig. 1(a)
where solid curves and marks denote
the results of the AMM and DS, respectively.
We note that for $J=0$, $\rho$ is inversely proportional
to $N$, as realized in Eq. (29).
With increasing $J$,
$\gamma$ is decreased while $\rho$ has no changes.
The results of AMM are in good agreement with
those of DS.

The $N$-dependences of
$\gamma$ and $\rho$ in the stationary states
for sigmoid couplings of $K=0.0$, 0.2 and 0.5
are plotted in Fig. 1(b) where solid curves denote
the result of the AMM [Eqs.(36) and (38)]
and where marks express
those of DS calculated by using $H(x)=x$.
With increasing $K$, $\rho$ is increased while
$\gamma$ is little changed except for $N < 5$. 

\subsubsection{Noise-strength $(\alpha)$ dependence}

The $\alpha$ dependences of stationary
$\gamma$ and $\rho$ for diffusive couplings
with $N=10$, $\lambda=1.0$ and $\beta=1.0$
are shown in
Figs. 2(a), where filled and open marks denote
$\gamma$ and $\rho$, respectively, in DS, and solid and chain curves
the respective results in the AMM.
Note that the results of $\rho$ are
multiplied by a factor of 10 $(=N)$, and that
three curves in the AMM are degenerated in Fig. 2(a):
$10 \rho(J=0.5) = 10 \rho(J=0.0) = \gamma(J=0.0)$.
For $J=0.0$, the relation of $\rho=\gamma/N$
holds in both the AMM and DS.
For $J=0.5$, $\gamma$ in the AMM is decreased 
compared to that for $J=0.0$, in agreement
with the result of DS.
In contrast, $\rho$ in the AMM is the same as
that of $J=0.0$
because it is independent of $J$ [Eq. (25)],
while $\rho$ in DS is decreased with increasing $J$
at $\alpha > 0.6$.
The results of the AMM diverge at $\alpha=1$,
where those of DS remain finite.
This difference in $\rho$ 
between the AMM and DS at large $\alpha$
is attributed to the second-moment approximation
in the AMM,
and it is the fallacy in the AMM
neglecting higher-order moments.

Figure 2(b) shows the $\alpha$-dependent $\gamma$
and $\rho$ for the sigmoid couplings of $K=0.5$.
Solid and chain curves denote $\gamma$ and $\rho$, respectively, 
in the AMM given by Eqs. (36) and (38): 
filled and open squares express those in DS with $H(x)=x$.
Although $\gamma$ and $\rho$ diverge at $\alpha=0.71$ in the AMM,
they persist up to $\alpha \sim 0.77$ in DS.

\subsubsection{Coupling $(J, K)$ dependence}

Figure 3(a) shows
the $J$ dependences of $\gamma$
and $\rho$ for the diffusive couplings
with $N=10$, $\alpha=0.5$ and $\beta=1.0$:
solid and chain curves express $\gamma$ and $\rho$, respectively,
in the AMM, and filled and open circle the respective results
in DS.
With increasing $J$, $\gamma$ is decreased while 
$\rho$ is almost independent of $J$.

The $K$ dependences of $\gamma$
and $\rho$ for the sigmoid couplings
are plotted in Fig. 3(b) where
solid and chain curves express $\gamma$ and $\rho$, respectively,
in the AMM, and filled and open circle 
denote $\gamma$ and $\rho$, respectively, in DS.
In the AMM, the critical coupling $K_c$ where 
$\gamma$ and $\rho$ diverge, is $K_c=0.75$
[Eqs.(36) and (38)] while DS leads to $K_c \sim 0.84$.

These differences realized in numerical
results are consistent with the items (ii) and (iii)
mentioned in Section 2.3.
We note in Figs. 3(a) and 3(b) that
the synchronization is increased with increasing $J$ and $K$
because Eq. (41) approximately yields
$S \propto (10 \rho - \gamma)$: $S$
is proportional to the difference between chain ($10 \rho$) 
and solid curves ($\gamma$).
This agrees with the item (iv) in Section 2.3 denoting
the similarity between the two couplings.

\subsection{Dynamical property}

We apply a pulse input given by
\begin{eqnarray}
I^{(e)}(t)= A\: \Theta(t-t_1) \Theta(t_2-t),
\end{eqnarray}
with $A=0.5$, $t_1=40$ and $t_2=50$,
where $\Theta(x)$ denotes the Heaviside function:
$\Theta(x)=1$ for $x > 0$ and zero otherwise.
Figures 4(a), 4(b) and 4(c) show the responses of
$\mu(t)$, $\gamma(t)$ and $\rho(t)$, respectively,
to the external input given by Eq. (46)
with $\alpha=0.5$, $\beta=1.0$, $J=0.0$ and $N=10$.
Solid curves denote the results of the AMM 
which are in good agreement with those of DS.
Input pulse induces changes not only in $\mu(t)$
but also in $\gamma(t)$ and $\rho(t)$.
These arise from terms of $\alpha^2\:\mu^2$ 
in Eqs. (24) and (25).
Indeed, in the case of $\alpha=0.0$,
input pulse induces no changes in $\gamma(t)$ and $\rho(t)$,
as shown by chain curves for the AMM result.

The response of $\mu(t)$ for the diffusive coupling 
is independent of the coupling $J$, as realized in Eq. (23).
In contrast, the response $\mu(t)$ for the sigmoid coupling
shows much variety depending on its coupling $K$.
Figure 5(a) and 5(b) show $\mu(t)$ and $S(t)$, respectively,
for various values of $K$ 
with $\alpha=0.5$, $\beta=0.0$ and $N=10$,
when the pulse input given by Eq. (46) is applied. 
With increasing $K$, magnitudes of $\mu(t)$
are increased at $40 < t < 50$ during which the input pulse
is applied.
It is interesting that the synchronization
$S(t)$ is decreased at $40 < t < 50$ by an applied input pulse
which reduces the ratio of $\rho/\gamma$,
and then $S(t)$ is much increased at $t > 50$.
For $K=0.0$, $S(t)$ vanishes because $\rho=\gamma/N$
in Eq. (41).
Figure 5(c) and 5(d) show similar
plots of $\mu(t)$ and $S(t)$, respectively,
for combined noises of $\alpha=0.5$ and $\beta=1.0$. 
With increasing $K$, the magnitude of $\mu(t)$ is again increased,
although an agreement between the results of the AMM and DSs
become worse than that shown in Fig. 5(a).
$S(t)$ is decreased by an applied pulse, but no increases
at $t > 50$, in contrast with the case shown in Fig. 5(b).

We have applied also the sinusoidal input given by
\begin{eqnarray}
I^{(e)}(t)= A \:\left[1-\cos \left(\frac{2 \pi t}{T_p} \right) \right],
\end{eqnarray}
where $A=0.5$ and $T_p=20$.
The responses of
$\mu(t)$, $\gamma(t)$ and $\rho(t)$ are shown in
Figs. 6(a), 6(b) and 6(c), respectively,
when the external input given by Eq. (46) is applied
for $\alpha=0.5$, $\beta=1.0$, $J=0.0$ and $N=10$.
Solid curves expressing the results of the AMM 
are in good agreement with dashed curves of those of DS.
Input pulse induces changes in $\mu(t)$ and also
in $\gamma(t)$ and $\rho(t)$.
For a comparison, we show, by chain curves, the AMM result
for $\alpha=0.0$, for which no changes in
$\gamma(t)$ and $\rho(t)$ by an applied input.


\section{DISCUSSION AND CONCLUSION}

It is interesting to discuss the stationary distribution
of our generalized Langevin model given 
by Eqs. (1)-(3). 
In the case of no couplings ($J=K=0$), 
the probability distribution
$\hat{p}(\{ x_i \},t)$ is given by
\begin{equation}
\hat{p}(\{ x_i \},t) = \Pi_i\: p(x_i, t),
\end{equation}
where $p(x_i, t)$ expresses the distribution for
a local variable $x_i$ satisfying the FPE given by
\begin{eqnarray}
\frac{\partial}{\partial t}\: p( x_i,t) 
&=&- \frac{\partial}{\partial x_i}\{ [F(x_i) 
+\frac{\phi\alpha^2}{2}G'(x_i)G(x_i) 
+ I^{(e)}] \:p( x_i,t)\}  
\nonumber \\
&+&\frac{1}{2} \frac{\partial^2}{\partial x_i^2} 
\{[\alpha^2 G(x_i)^2+\beta^2]\:p(x_i,t) \}.
\end{eqnarray}
For a constant input of $I^{(e)}(t)=I$,
the stationary distribution $p(x_i)$
is expressed by \cite{Schenzle79}-\cite{Ante05}
\begin{eqnarray}
\ln p(x) &=& X(x)+Y(x)
-\left(1- \frac{\phi}{2} \right)
\ln \left[\frac{\alpha^2 G(x)^2}{2}+\frac{\beta^2}{2} \right],
\end{eqnarray}
with
\begin{eqnarray}
X(x) &=& 2 \int \:dx \:
\left[ \frac{F(x)}{\alpha^2 G(x)^2+\beta^2} \right], \\
Y(x) &=& 2 \int \:dr \:
\left[ \frac{I}{\alpha^2 G(x)^2+\beta^2} \right].
\end{eqnarray}

For the linear Langevin model with $F(x)=-\lambda x$ and $G(x)=x$, 
$p(x)$ in the Stratonivich representation becomes
\begin{eqnarray}
p(x) &\propto& 
\left[1+\left( \frac{\alpha^2}{\beta^2} \right)x^2 
\right]^{-(\lambda/\alpha^2+1/2)}
\exp[Y(x)], 
\end{eqnarray}
with
\begin{eqnarray}
Y(x)=\left( \frac{2 I}{\alpha \beta} \right)
{\rm arctan}\left( \frac{\alpha x}{\beta} \right).
\end{eqnarray}
We examine the some limiting cases of 
Eq. (53) as follows.

\noindent
(a) Equation (53) in the case of $I=Y(x)=0$ 
expresses the $q$-Gaussian
\cite{Sakaguchi01,Anten02,Tsallis88,Tsallis98},
which becomes, in the limit of large $x$ ($\gg \beta/\alpha$),
\begin{equation}
p(x) \propto x^{-\delta},
\end{equation}
with
\begin{equation}
\delta= \frac{2 \lambda}{\alpha^2}+1.
\end{equation}
The expectation value of $x^2$ is given by
\begin{equation}
\langle x^2 \rangle = \frac{\beta^2}{2(\lambda-\alpha^2)},
\end{equation}
which requires $\alpha^2 < \lambda$ for positive definite 
$\langle x^2 \rangle$.

\noindent
(b) For $\alpha=0$ and $\beta \neq 0$, we get from Eq. (53)
\begin{eqnarray}
p(x) &\propto& \exp\left[ -\left(\frac{\lambda}{\beta^2} \right)
\left( x- \frac{I}{\lambda} \right)^2 \right]. 
\end{eqnarray}

\noindent
(c) For $\beta=0$ and $\alpha \neq 0$, Eq. (53) becomes
\begin{eqnarray}
p(x) &\propto& x^{-(2\lambda/ \alpha^2+1)}
\exp\left[ -\left(\frac{2I}{\alpha^2} \right)\frac{1}{x} \right].
\end{eqnarray}

Figures 7(a)-7(c) show the distribution $p(x)$
calculated with the use of Eqs. (53)-(59).
The distribution $p(x)$ for $\alpha=0.0$ in Fig. 7(a)
shows the Gaussian distribution given by Eq. (58)
without multiplicative noises,
which is shifted by an applied input $I$.
When multiplicative noises are added ($\alpha \neq 0$), 
the form of $p(x)$
is changed but the average of $ \langle x \rangle $ is conserved
as shown in Fig. 7(a). 
Figure 7(b) shows that when the magnitude of additive noises
$\beta$ is increased,
the width of $p(x)$ is increased.
We note in Fig. 7(c) that
although $p(x)$ is symmetric for $I=0$, 
the external input $I$ increases
the asymmetry in $p(x)$. 
Figures 7(a)-7(c) clearly show that
$p(x)$ is much modified by the presence of $I$.


Now we consider the averaged, global variable $X(t)$
given by Eq. (8).
The stationary distribution for a global variable $X$
given by Eq. (9), is analytically expressed
only for limited cases.

\noindent
(a) For $\beta \neq 0$ and $ \alpha=0$, $P(X)$ is given by
\begin{eqnarray}
P(X) &\propto& 
\exp\left[ -\left( \frac{\lambda N}{\beta^2} \right)
\left( X- \frac{I}{\lambda} \right)^2 \right], 
\end{eqnarray}
which arises from the central-limit theorem for $\beta$.

\noindent
(b) For $I=0$, we get
\begin{equation}
P(X)=\frac{1}{2 \pi} \int_{-\infty}^{\infty}\: dk
\;e^{i k X}\:\Phi(k),
\end{equation}
with
\begin{equation}
\Phi(k)=\left[\phi\left( \frac{k}{N} \right) \right]^N,
\end{equation}
where $\phi(k)$ is the characteristic function
for $p(x)$ given by \cite{Abe00}
\begin{eqnarray}
\phi(k)&=& \int_{-\infty}^{\infty} \;
e^{-i k x}\:p(x) dx, \\
&=& 2^{1-\nu}\frac{(\lambda' \mid k \mid )^{\nu}}{\Gamma(\nu)}
K_{\nu}(\lambda' \mid k \mid),
\end{eqnarray}
with
\begin{eqnarray}
\nu&=& \frac{\lambda}{\alpha^2}, \\
\lambda'&=& \frac{\beta}{\alpha}, 
\end{eqnarray}
$K_{\nu}(x)$ expressing the modified Bessel
function.

The asymptotic form of $P(X)$ for large $X$ and large $N$
is obtained as follows.
By using the relation:
\begin{equation}
z^{\nu} B_{\nu}(z) \propto \left[1 - c z^2 - d z^{2 \nu}
+ \cdot \cdot \right],
\hspace{1cm}\mbox{for $z \ll 1$, $\nu \neq 1$}
\end{equation} 
with 
\begin{eqnarray}
c&=& \frac{1}{4(\nu-1)}, \\
d&=& \left( \frac{1}{2} \right)^{2 \nu}
\frac{\Gamma(1-\nu)}{\Gamma(1+\nu)},
\end{eqnarray}
we get, for large $N$, 
\begin{eqnarray}
\Phi(k) &\propto& \exp(-a_N \:k^{2}),
\hspace{1.5cm}\mbox{for $\nu > 1$} \\
&\propto& \exp(-b_N \mid k \mid^{2\nu} ),
\hspace{1cm}\mbox{for $0 < \nu < 1$}
\end{eqnarray}
with
\begin{eqnarray}
a_N &=& c N^{-1} (\lambda')^2, \\
b_N &=& d N^{1-2\nu} (\lambda')^{2 \nu}.
\end{eqnarray}
For large $X$, Eqs. (61), (70)-(73) yield
\begin{eqnarray}
P(X) &\propto& \exp\left( -\frac{X^2}{2 \sigma_N^2} \right),
\hspace{1cm}\mbox{for $\nu > 1$} \\
&\propto& X^{-\delta'},
\hspace{2.5cm}\mbox{for $0 < \nu < 1$}
\end{eqnarray}
with
\begin{eqnarray}
\sigma_N^2&=&2 a_N = \frac{\beta^2}{2N(\lambda-\alpha^2)}, \\
\delta'&=&2 \nu+1= \frac{2 \lambda}{\alpha^2}+1.
\end{eqnarray}
It is interesting that for $N=1$, Eq. (76) coincides
with Eq. (57) and the index of $\delta'$ given by Eq. (77) is
the same as $\delta$ given by Eq. (56).
The case of $\nu=1$, excluded in the above analysis,
will be numerically studied below.
The stable distribution of $P(X')$ 
for $X'(t)= N^{-1/2 \nu} \sum_i x_i(t)$
with $0 < \nu < 1$ was discussed in Ref. \cite{Abe00}. 

Figures 8(a) shows distributions of
a global variable $P(X)$ for $I=0.0$ calculated by
DS for the Langevin model
given by Eq. (1) with $N=1$, $N=10$ and $N=100$
($\lambda=1.0$, $\alpha=1.0$, $\beta=0.5$ and $\nu=1.0$).
For a comparison, 
results of the analytic expression given by 
Eqs. (61), (62) and (64) are plotted 
with a shift by $X=-2$ for a clarity of the figure.
We note that with increasing $N$,
the width of $P(X)$ becomes narrower,
which is consistent with the central-limit theorem.
Figures 8(b) and 8(c) show $P(X)$ 
for $I=1.0$ and $I=2.0$, respectively,
calculated by DS.
The $N$ dependence of $P(X)$ for finite $I$ is intrigue:
with increasing $N$,
not only its width becomes narrower
but also its profile is considerably modified,
as shown in Figs. 8(b) and 8(c).
This trend is more significant for a larger $I$. 

So far we have assumed the vanishing couplings,
which is now introduced.
Figure 9(a) shows distributions of $p(x)$ and $P(X)$
for the diffusive couplings of 
$J=0.0$ (dashed curves) and $J=1.0$ (solid curves)
with $I=0.0$.
We note that with increasing $J$, the width of $p(x)$
becomes narrower while that of $P(X)$ is not changed.
This is the case also for finite $I=1.0$, as shown in Fig. 9(b).

Figure 9(c) shows $p(x)$ and $P(X)$ 
with $I=0$ for the sigmoid coupling of $K=0.0$ (dashed curve)
and $K=0.5$ (solid curve).
We note that the width of $P(X)$ for $K=0.5$ become wider
than that for $K=0.0$. 
Figure 9(d) shows that an introduction
of $K$ with finite $I=1.0$ induces not only an increase
in the width of $P(X)$ but also its shift.
This is in contrast with the case of the diffusive 
coupling shown in Fig. 9(b), where $P(X)$ has little 
effects of $J$.

The coupling dependences 
of stationary distributions of 
$p(x)$ and $P(X)$
are related to those of
$\gamma$ and $\rho$,
because $\sqrt{\gamma}$ and $\sqrt{\rho}$
approximately express the widths of $p(x)$ and $P(X)$,
respectively.
Figure 9(a) shows that with increasing $J$, 
the width of $p(x)$ is decreased
while that of $P(X)$ is unchanged for the diffusive couplings.
In contrast, for the sigmoid coupling,
the width of $P(X)$ is increased 
while that of $p(x)$ is unchanged when $K$ is increased,
as shown in Fig. 9(c).
These are consistent with 
the dependences of $\gamma$ and $\rho$ on the 
type of couplings expressed in
the items (ii) and (iii)
having been discussed in Section 2.3.

\section{CONCLUSION}

By using the AMM, we have studied
stationary and dynamical properties of finite $N$-unit
Langevin model which is subjected to multiplicative noises
and which is coupled by diffusive 
and sigmoid couplings.
Properties of coupled Langevin model are shown to
depend on both the type and magnitude
of couplings, which is the same
as in the case of FitzHugh-Nagumo model 
\cite{Hasegawa05a}\cite{Hasegawa06b}.
One of advantages of the AMM is that we may easily solve
the low-dimensional differential equations 
although its applicability
is limited to the weak-noise case. 
It would be interesting to apply
the AMM to various types of stochastic 
coupled ensembles such as neuronal and complex networks
in order to discuss their dynamics.

\section*{Acknowledgements}
This work is partly supported by
a Grant-in-Aid for Scientific Research from the Japanese 
Ministry of Education, Culture, Sports, Science and Technology.  




\newpage

\begin{figure}
\caption{
(Color online)
(a) The $N$ dependences of the stationary 
$\gamma$ and $\rho$
for various diffusive couplings (DC):
circles, triangles and square denote results of DS
for $J=0.0$, $J=0.2$ and $J=0.5$, respectively, and 
solid curves express those of the AMM. 
(b) The $N$ dependence of the stationary 
$\gamma$ and $\rho$
for various sigmoid couplings (SC)
with $H(x)=x$:
circles, triangles and square denote results of DS
for $K=0.0$, $K=0.2$ and $K=0.5$, respectively,
solid curves express those of the AMM 
($\lambda=1.0$, $ \alpha=0.5$ and $ \beta=1.0$).
Dashed curves are drawn only for a guide of the eye.
}
\label{fig1}
\end{figure}

\begin{figure}
\caption{
(Color online)
(a) The $\alpha$ dependences of the stationary 
$\gamma$ (solid curves) and $\rho$ (chain curves)
for the diffusive couplings (DC):
circles, triangles and square denote results of DS
for $J=0.0$ and $J=0.5$, respectively, and 
solid and chain curves express those of the AMM. 
(b) The $\alpha$ dependence of the stationary 
$\gamma$ (solid curves) and $\rho$ (chain curves)
for sigmoid couplings (SC)
of $K=0.5$ with $H(x)=x$:
squares denote results of DS, and
solid and chain curves express those of the AMM.
($N=10$, $\lambda=1.0$ and $ \beta=1.0$).
Dashed curves are drawn only for a guide of the eye.
}
\label{fig2}
\end{figure}

\begin{figure}
\caption{
(Color online)
(a) The $J$ dependences of the stationary 
$\gamma$ (solid curves) and $\rho$ (chain curves)
for the diffusive couplings (DC):
circles denote results of DS, and 
solid and chain curves express those of the AMM. 
(b) The $K$ dependence of the stationary 
$\gamma$ (solid curves) and $\rho$ (chain curves)
for sigmoid couplings (SC)
with $H(x)=x$:
squares denote results of DS, and
solid and chain curves express those of the AMM. 
($N=10$, $\lambda=1.0$, $\alpha=0.5$ and $ \beta=1.0$).
Dashed curves are drawn only for a guide of the eye.
}
\label{fig3}
\end{figure}

\begin{figure}
\caption{
(Color online)
Responses of 
(a) $\mu(t)$, (b) $\gamma(t)$ and (c) $\rho(t)$
to the pulse input with
$J=0.0$, $N=10$, $\lambda=1.0$, and $\beta=1.0$:
solid and dashed curve denote results of AMM and DS,
respectively, for $\alpha=0.5$, and chain curves 
express that of AMM for $\alpha=0.0$.
}
\label{fig4}
\end{figure}

\begin{figure}
\caption{
(Color online)
(a) Responses of $\mu(t)$ and (b) $S(t)$
to the pulse input for various sigmoid couplings
with $\lambda=1.0$, $\alpha=0.5$, $\beta=0.0$ and $N=10$.
(c) Responses of $\mu(t)$ and (d) $S(t)$
to the pulse input for various sigmoid couplings
with $\lambda=1.0$, $\alpha=0.5$, $\beta=1.0$ and $N=10$.
Solid and dashed curves denote results of AMM and DS,
respectively.
}
\label{fig5}
\end{figure}

\begin{figure}
\caption{
(Color online)
Responses of 
(a) $\mu(t)$, (b) $\gamma(t)$ and (c) $\rho(t)$
to the sinusoidal input with
$J=0.0$, $N=10$, $\lambda=1.0$, and $\beta=1.0$:
solid and dashed curve denote results of AMM and DS,
respectively, for $\alpha=0.5$, and chain curves 
express that of AMM for $\alpha=0.0$.
}
\label{fig6}
\end{figure}

\begin{figure}
\caption{
(a) Distributions $p(x)$ of local variable $x$
for various $\alpha$ with $\lambda=1.0$, 
$\beta=1.0$ and $I=1.0$,
(b) $p(x)$
for various $\beta$ with $\lambda=1.0$, 
$\alpha=1.0$ and $I=1.0$, and
(c) $p(x)$
for various $I$ with $\lambda=1.0$, 
$\alpha=1.0$ and $\beta=0.5$.
}
\label{fig7}
\end{figure}

\begin{figure}
\caption{
(Color online)
Distributions $P(X)$ of global variable $X$
calculated by direct simulation (DS)
for $N=1$ (dashed curves), $N=10$ (solid curves)
and $N=100$ (chain curves) 
with (a) $I=0$, (b) $I=1.0$ and (c) $I=2.0$
($\lambda=1.0$, $\alpha=1.0$ and $\beta=0.5$).
Results calculated with the use of Eqs. (67), (68) and (70)
are plotted in (a) with a shift of $X=-2$ 
for a clarity of the figure.
}
\label{fig8}
\end{figure}

\begin{figure}
\caption{
(Color online)
Distributions of $p(x)$ and $P(X)$
of local and global variables, respectively,
with (a) $I=0.0$ and (b) $I=1.0$
for diffusive coupling (DC)
with $J=0.0$ (dashed curves) 
and $J=1.0$ (solid curves), and
those with (c) $I=0.0$ and $I=1.0$
for sigmoid coupling (SC)
with $K=0.0$ (dashed curves)
and $K=0.5$ (solid curves).
($N=10$, $\lambda=1.0$, $\alpha=0.5$ and
$\beta=1.0$).
}
\label{fig9}
\end{figure}

\end{document}